\documentclass[twocolumn]{aastex62}

\usepackage{extarrows} 
\usepackage{isodateo}
\usepackage{graphicx}
\usepackage{xcolor}
\usepackage[utf8]{inputenc}
\renewcommand{\FR}[2]{\displaystyle\frac{\,{#1}\,}{#2}}
\newcommand{\fr}[2]{\mbox{$\frac{\,{#1}\,}{#2}$}}
\newcommand{\n}{\nonumber}

\graphicspath{{fig/}}

\newcommand{\bge}{\begin{equation}}
\newcommand{\ede}{\end{equation}}
\newcommand{\bga}{\begin{aligned}}
\newcommand{\eda}{\end{aligned}}
\newcommand{\bgp}{\begin{pmatrix}}
\newcommand{\edp}{\end{pmatrix}}
\newcommand{\bgs}{\begin{subequations}}
\newcommand{\eds}{\end{subequations}}
\newcommand{\order}[1]{\mathcal{O}({#1})}
\newcommand{\di}{{\mathrm{d}}}

\newcommand{\pd}{\partial} 

\renewcommand{\to}{\rightarrow}

\renewcommand{\ga}{\gamma}

\newcommand{\lam}{\lambda}
\newcommand{\rh}{\rho}

\newcommand{\wh}[1]{\mkern 2mu \widehat{\mkern-2mu#1\mkern-2mu}\mkern 2mu}
 
\newcommand*{\vcenteredhbox}[1]{\begingroup
\setbox0=\hbox{#1}\parbox{\wd0}{\box0}\endgroup}

\begin{document} 

\title{A Direct Probe of Mass Density Near Inspiraling Binary Black Holes}

\author{Lisa Randall}
\affiliation{Department of Physics, Harvard University, 17 Oxford St., Cambridge, MA 02138, USA} 

\author{Zhong-Zhi Xianyu}
\affiliation{Department of Physics, Harvard University, 17 Oxford St., Cambridge, MA 02138, USA}  
 
\begin{abstract}
Now that LIGO has revealed the existence of a large number of binary black holes, identifying their origin emerges as an important challenge. Observed binary black holes might reside in more isolated regions of the galaxy or alternatively they might be emerging from  dense environments such as galactic centers or globular clusters. In the latter case, their center of mass motion as well as their orbital parameters lead to potentially observable changes in the waveforms that would reflect their gravitational interactions with the surrounding matter. The gravitational wave signal would be distinguished by a net phase change or even a time-dependent Doppler shift arising from the orbital motion. We show that this time-dependence might be observable in future space gravitational wave detectors such as LISA that could provide direct information about the black hole binary environments and otherwise invisible ambient mass.
 
\vspace{1.5cm}
\end{abstract}

\section{Introduction}
 
The LIGO detection \cite{Abbott:2016blz} of gravitational waves (GWs) from a pair of inspiraling black holes (BHs) has launched a new era of GW astronomy. BBHs could show up in the LISA band months before they merge in the LIGO band and a joint detection from both frequency bands would allow for better measurements of their properties \cite{Sesana:2016ljz}. 

The LISA observation of stellar-mass BBHs should be particularly useful in understanding the formation channel of BBHs. Current strategies for differentiating BBH formation channels focus on measuring the statistical distribution of orbital parameters, either eccentricity or spin. BBHs formed in isolation generally have circular orbits and aligned spins, while dynamically formed BBHs could possess large spin misalignment and finite eccentricity. The nonzero eccentricity is a generic feature of dynamical channels, which can be generated  due to the ``thermal'' distribution of eccentricity $f(e)\di e\propto e\di e$ in dense environments, or via Kozai-Lidov resonances with a third body, or alternatively from non-perturbative multi-body processes.
In this paper we will show that these channels lead to different eccentricity distributions so a careful measurement of eccentricity will be useful not only to distinguish between the isolated and dynamical channels, but also to further differentiate among different modes of dynamical formation. Such a powerful measurement of  eccentricity distributions can be done  only with low frequency observations such as those proposed for LISA \cite{Nishizawa:2016jji,Nishizawa:2016eza,Breivik:2016ddj,Rodriguez:2016vmx,Randall:2017jop,Randall:2018nud,Banerjee:2017mgr,Samsing:2018isx,Kremer:2018cir}.

In this paper, we point out a more ambitious way of studying BBHs, building on a special property of LISA measurements, namely the capacity for measuring directly the barycenter motion of BBHs orbiting around nearby objects and the corresponding orbital elements. Since the time dependence of the phase reflects the orbital period of  the barycenter motion  $T\sim2\pi(Gm/r^3)^{-1/2}$, the time scale for the phase variation in the case of circular outer orbits serves as a direct probe of the ambient density $\rh\sim m/r^3$, where $m$ is the mass enclosed by the orbits and $r$ is the distance over which the BBH orbits. Even when not circular, this measurement probes the matter distribution but in a more subtle way  \cite{2019arXiv190201344H,2019arXiv190201345H}. For  favorably positioned BBHs, the detailed barycenter motion in the LISA band could also be measurable, raising a unique opportunity to directly measure the orbital elements of the two-body system formed by the BBH and a nearby tertiary mass. 

This way of viewing the LISA signal, for which there will be many well-determined GW phases with a slower time-dependence that reflects external interactions, is analogous to measurements of a pulsar in a binary. 
As with pulsars, this measurement depends primarily on phase measurements. Employing also a well-measured amplitude will provide a complementary way of increasing precision  and checking consistency
with
 the phase measurement. Furthermore, a sufficiently accurate measurement of the orbital motion can allow us to follow the waveform in detail  from the LISA to the LIGO window.

Previous studies have considered the effect of barycenter acceleration arising from cosmological inhomogeneity and local gravity in different parameter regions \cite{Bonvin:2016qxr,Yunes:2010sm,Meiron:2016ipr}.   Because the net phase shift can be confused with a chirp signal, a Fisher matrix analysis is critical. We consider all possible third-body mass from stellar-mass objects to supermassive BHs, and extend this analysis to include the cross-correlation between the outer and inner binary parameters to clarify further which parameter range leads to a detectable orbit. We also consider more directly observing the barycenter motion of the BBH orbiting around a nearby massive object
through direct measurements of the time-dependence of the phasing owing to the BBH motion, similar to the analysis in \cite{Robson:2018svj} which considered astrophysical objects with little chirping.  We extend the analysis to chirping BBHs which are important targets for multi-band observations, and use Fisher matrix to quantify possible confusions. We also extend the analysis to allow for the
secular variation of the inner orbital elements, including the eccentricity and the inclination due to the Kozai-Lidov (KL) mechanism \cite{Kozai:1962zz,Lidov:1976qhg}.

\section{Motion of the BBH barycenter along the Outer Orbit}

We will call the effective two-body system formed by the  BBH barycenter and the tertiary mass the \emph{outer binary} and its orbit the \emph{outer orbit} denoted with subscript ``2", and call the BBH emitting GWs in LISA the \emph{inner binary} with subscript ``1".   The barycenter motion of the BBH would introduce a time variation of the apparent frequency via the time-varying Doppler shift $\Delta z(t)$ of the GW signals.

  We consider a pair of BHs with mass $m_0,m_1\sim\order{10M_\odot}$. LISA  could access such a pair if the GW peak frequency $f_\text{peak}$ falls roughly between $0.01$Hz and $0.1$Hz, where $f_\text{peak}$ is
\bge
\label{fpeak}
  f_\text{peak}(t)=\FR{\sqrt{Gm}}{\pi\big[a_1(t)\big(1-e_1^2(t)\big)\big]^{3/2}}\big(1+e_1(t)\big)^{1.1954},
\ede
where $G$ is Newton's constant, $a_1$ and $e_1$ denote the semi-major axis and the eccentricity of the inner binary, and $m=m_0+m_1$ is the total mass of the BBH.  
As we shall show below, the time-dependence of the long-term orbital motion  and/or the secular change of inner eccentricity has a time scale comparable or much longer than a year in general. Therefore these effects will be visible only if the BBH stays long enough ($\gtrsim 1$yr) in the LISA band. 

We model the ambient density of a BBH by a single tertiary body. This model can approximate  1) a BBH orbiting around an SMBH in the galactic center \cite{Antonini:2012ad,Randall:2017jop,Randall:2018nud}; 2) a BBH orbiting around an IMBH in a globular cluster; 3) a BBH orbiting stellar mass in a nuclear cluster;
4) a BBH in stellar-mass hierarchical triples in globular clusters \cite{Wen:2002km} or in the field \cite{Silsbee:2016djf}. The effective mass $m_2$ of the tertiary body can be in the range $\order{10M_\odot}$ to $\order{10^9M_\odot}$ or even larger.

The effect of the outer orbital motion is easy to see 
with Newtonian dynamics
 for a circular outer orbit and constant GW frequency $f_\text{peak}$, in which case the GW phase is $\Phi(t_\text{BBH})=f_\text{peak}t_\text{BBH}$ in the BBH frame where $t_\text{BBH}$ is the time defined in the source frame. The time of arrival $t$  is then  related to the time in the source frame $t_\text{BBH}$ by $t=t_\text{BBH}+r_{2\parallel}/c$. Here $r_{2\parallel}$ is the location of the BBH barycenter along the line of sight, and is given by $r_{2\parallel}=(m_2/M)a_2\sin(2\pi t/P_2)$ in this simple case, where $a_2$ and $P_2$ are the  semi-major axis and the orbital period of the outer orbit and are related by $2\pi P_2^{-1}=\sqrt{GM/a_2^3}$, with $M=m_0+m_1+m_2$. Therefore,
\bge
\label{phisimple}
 \Phi(t)= f_\text{peak}  \Big(t-\FR{m_2}{M}\FR{a_2}{c} \sin\FR{2\pi t}{P_2}\Big).
\ede
This phase dependence 
assumes small peculiar acceleration of the triple barycenter.
In reality, $f_\text{peak}(t)$ also depends on time $t$ through its dependence on $a_1(t)$ and $e_1(t)$ as shown in (\ref{fpeak}).  
The linear time-dependence in (\ref{phisimple}) is indistinguishable from a redshift attributable to a different distance. For the effect to be unambiguously distinguishable, either the time has to be long enough for all terms in a Taylor expansion to contribute, or we need to otherwise subtract off the inner orbital contribution to the  $\order{t^2}$ term.

The first condition states that we need to follow the phase over a time comparable to the orbital period $P_2$ (ideally less than half an orbit).

In order to evaluate  the second condition, we  include the full time dependence of both the inner and outer orbits. To see the time shift entering the GW phase from $r_{2\parallel}$, we note that in the BBH barycenter frame, the GW phase $\Phi$ is,
\bge
\label{phidef}
  \Phi(t_\text{BBH})=\int_0^{t_\text{BBH}} f_\text{peak}(t_\text{BBH}')\di t_\text{BBH}',
\ede   
where we have dropped a constant phase by shifting the initial point of $t_\text{BBH}$.  Replacing $t_\text{BBH}$ in (\ref{phidef}) by $t$, we find,
\begin{align}
\label{phiobs}
   \Phi(t)=
&\int_0^{t} 
f_\text{peak}(t')\bigg[1-\FR{v_{2\parallel}(t')}{c}\bigg]\di t' , 
\end{align}
where $v_{2\parallel}=\dot r_{2\parallel}$.

Choosing the plane orthogonal to the line of sight as the reference plane, the time of arrival of GW phases will be delayed by $r_{2\parallel}/c$, where $r_{2\parallel}$ is the barycenter position along the line of sight. In the barycenter system of the triple, $r_{2\parallel}$ can be represented in terms of orbital elements as
\bge
\label{r2p}
r_{2\parallel}=\FR{m_2}{M}\FR{a_2(1-e_2^2)}{1+e_2\cos\psi_2}\sin(\ga_2+\psi_2)\sin I_2,
\ede
where $e_2$, $I_2$, $\ga_2$, and	 $\psi_2$ are the outer orbital elements, corresponding to the eccentricity, the inclination, the argument of the periapsis, and the true anomaly, respectively. The factor $m_2/M$ is from the conversion to the barycenter frame of the triple from the reduced coordinates. Note that this means that the BBH motion relative to the triple barycenter is suppressed if $m_2\ll M$ and will be small unless the tertiary body is comparable to  or greater in mass than those in the binary system.

\begin{figure}[t]
\centering
\includegraphics[width=0.45\textwidth]{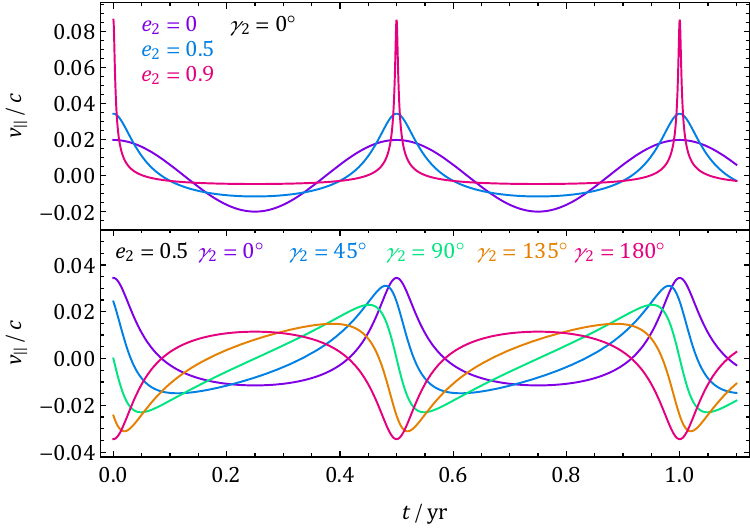}
\caption{Longitudinal velocity $v_{2\parallel}(t)$ of the BBH barycenter. In both panels $m_0=m_1=10M_\odot$, $m_2=4\times 10^6M_\odot$ (the mass of Sgr A*), $a_2=100$AU, $I_2=90^\circ$. Note that the inclination $I_2=90^\circ$ optimizes the effect so better velocity sensitivity will be important in general cases. \label{fig_voft}}
\end{figure}

We see from (\ref{phiobs}) that the net effect of the time delay is a shift of GW frequency $f_\text{peak}\to (1-v_{2\parallel}/c)f_\text{peak}$, which is just the familiar (longitudinal) Doppler shift, where we neglect the smaller transverse component. We also neglect the suppressed relativistic corrections including the Roemer delay and Shapiro delay. 
We show in Fig.\;\ref{fig_voft} the fluctuation of $v_{2\parallel}/c$ over two periods of outer orbital motion for parameters described in the caption. 
 
\section{Observability}
Now we study the observability of the barycenter motion described in the last section, with a simplified Fisher matrix analysis. 
Instead of including all parameters characterizing the inner and the outer orbits, we pay special attention to the following three parameters: the binary's chirp mass $m_c\equiv (m_0m_1)^{3/5}/m^{1/5}$, the outer orbital frequency $\Omega\equiv 2\pi/P_2$, the maximal variation of the binary barycenter's transverse velocity $u$ within one outer-orbital period. For circular outer orbit we have $\Omega=\sqrt{GM/a_2^3}$ and  $u = 2(m_2/M)\sqrt{GM/a_2}\sin I_2$. By examining $\Omega$ and $u$ we hope to learn how well the outer orbit can be resolved, and by studying the correlation between $m_c$ and $\Omega$ we will demonstrate to what extent we can possibly remove the confusion between the binary chirping and the outer orbital motion.

For a GW waveform $h(t)$ in time domain, we follow \cite{Barack:2003fp} and define the corresponding noise-weighted waveform $\wh h(t)=h(t)/S_N^{1/2}(f(t))$, where $S_N(f)$ is the noise strain of LISA, and $f(t)$ is the frequency of the waveform $h(t)$ at time $t$. For GW with more than one harmonic component, the definition should be generalized to $\wh h_n(t)=h_n(t)/S_N^{1/2}(f_n(t))$, with $h_n$ the $n$'th harmonic component of the waveform and $f_n$ its frequency. Then, the signal-to-noise ratio (S/N) can be calculated by
\bge
\label{SNR}
  (S/N)^2=2\sum_n\int_0^{T_O}\di t\,\wh h_n^2(t),
\ede
where $T_O$ is the time of observation.

\begin{figure}[t]
\centering
\includegraphics[width=0.4\textwidth]{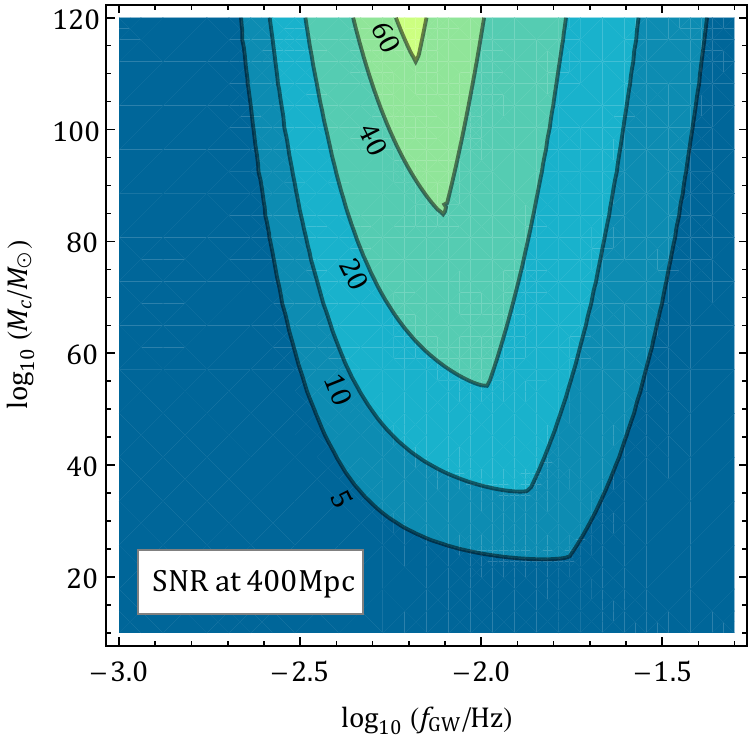}
\caption{The LISA S/N of a circular binary with the chirp mass $m_c$ and initial GW frequency $f_{GW}$, at a distance of $r=400$Mpc, with N2A5 configuration and $T_O=5$yr.}
\label{fig_SNR}
\end{figure}

In Fig.\;\ref{fig_SNR} we show the S/N for a circular binary at a distance of 400Mpc over a range of relevant masses and frequencies, assuming a five-year observation time of LISA, with noise curve with N2A5 configuration taken from \cite{Klein:2015hvg}. We note that the S/N here is calculated assuming the match filtering with appropriate templates that allow for BBH orbital motion. If the outer orbital parameters are properly taken into account by the templates, the resulting S/N is essentially independent of these parameters, because the S/N is mainly controlled by the power of the GW radiation while the outer orbital parameters has little influence on the GW power of the binary.

In addition, let $\lam_i~(i=1,2,\cdots)$ represent the parameters on which the waveform $h(t)$ depends. Then the error $\Delta \lam_i$ of measuring $\lam_i$ and the error correlation $\rh_{\lam_i\lam_j}$ between $\lam_i$ and $\lam_j$ are given respectively by, 
\begin{align}
  &\Delta \lam_i=\Sigma^{1/2}_{\lam_i\lam_i},
  &&\rh_{\lam_i\lam_j}^2= \Sigma_{\lam_i\lam_j}^2/\Sigma_{\lam_i\lam_i}\Sigma_{\lam_j\lam_j} .
\end{align}
Here the matrix $\Sigma$ is the inverse of the Fisher matrix $\Gamma$, which can be calculated by
\bge
\label{FisherM}
  \Gamma_{\lam_i\lam_j}=2\sum_n\int_0^{T_O}\di t\,\pd_{\lam_i}\wh h_n(t)\pd_{\lam_j}\wh h_n(t).
\ede
(\ref{SNR}) and (\ref{FisherM}) are readily used for a numerical study of the S/N and the parameter estimation. Here we will derive analytical approximations for several parameters, in order to develop physical intuition to more efficiently evaluate the potential range of parameters to be explored.

For simplicity, we consider binaries with circular orbits. We also  neglect post-Newtonian corrections and therefore restrict attention to the quadrupole GW. Then the waveform is single-frequency and can be written as $h(t)=h_c(t)\sin\Phi(t)$, where $h_c(t)$ is the signal's characteristic amplitude and $\Phi(t)$ its phase, given by (\ref{phiobs}). Generally, we have $\pd_{\lam_i}h=(\pd_{\lam_i}h_c)\sin\Phi+h_c(\pd_{\lam_i}\Phi)\cos\Phi\simeq h_c(\pd_{\lam_i}\Phi)\cos\Phi$. The last approximation holds because the total number of phases $\Phi\sim f_\text{GW}T_O\gg 1$ over a long period of observation. This is important in that it argues that the dominant contribution to the Fisher matrix error estimation arises from the phase only whereas the $S/N$ will depend on amplitude, provided that the templates are available.

When the error correlations are small, i.e., $\rh_{\lam_i \lam_j}^2\ll 1$ for all $i\neq j$, we have $\Sigma_{\lam_i\lam_i}\simeq (\Gamma_{\lam_i\lam_i})^{-1}$. In this case,  we have approximately,
\begin{align}
\label{DeltaLambda}
  \Delta{\lam_i}\simeq &~\Gamma_{\lam_i\lam_i}^{-1/2}\simeq 2\bigg[\int_0^{T_O}\di t\,\FR{(h_c\cos\Phi)^2(\pd_{\lam_i}\Phi)^2}{S_N}\bigg]^{-1/2}\n\\
  \simeq&~(S/N)^{-1}\big|\pd_{\lam_i}\Phi(T_O)\big|^{-1},
\end{align}
which makes sense. We see that larger $S/N$ and more sensitivity on the parameter $\lam_i$ (hence larger $\pd_{\lam_i}\Phi$) both help to reduce the error $\Delta\lam_i$.

Applying ($\ref{DeltaLambda}$) to a circular outer orbit, the two important parameters for which we want to evaluate sensitivity are the outer orbital frequency $\Omega $ and the amplitude of the transverse velocity $u $. 

Taking $\lam_i$ in (\ref{DeltaLambda}) to be $\Omega$ and $u$, respectively, we get the corresponding relative errors of parameter fitting,
\begin{align}
\label{DeltaOmegau}
  &\FR{\Delta\Omega}{\Omega}\simeq\FR{1}{(S/N)|\Phi_\Omega(T_O)|},
  ~~\FR{\Delta u}{u}\simeq\FR{1}{(S/N)|\Phi_u(T_O)|},\n\\
  &\Phi_\Omega(t)\equiv 2\pi\Omega\FR{u}{c} \int_0^{t}\di t' \,f_{GW}(t')\cos(\Omega t'+\ga_2)t',\n\\
  &\Phi_u(t)\equiv 
2\pi \FR{u}{c}\int_0^{t}\di t' \,f_{GW}(t')\sin(\Omega t'+\ga_2).
\end{align}
For slowly varying $f_{GW}(t)$, it is clear that $S/N\propto T_O^{1/2}$ and the measurement errors scale with observation time $T_O$ as $\Delta\Omega/\Omega\propto T_O^{-5/2}$ and $\Delta u/u\propto T_O^{-3/2}$, respectively.

\begin{figure*}[t]
\centering
\vcenteredhbox{\includegraphics[height=0.32\textwidth]{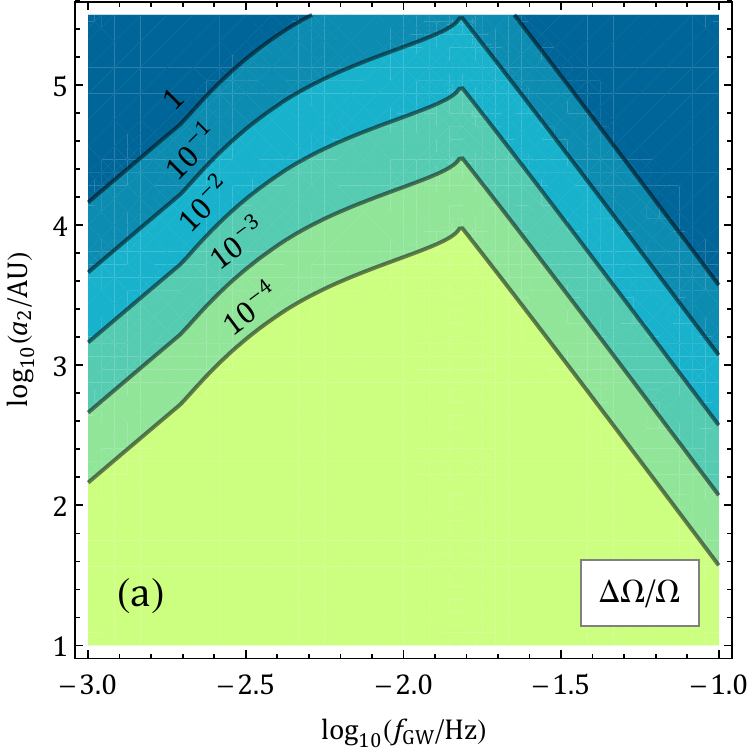}}
\vcenteredhbox{\includegraphics[height=0.32\textwidth]{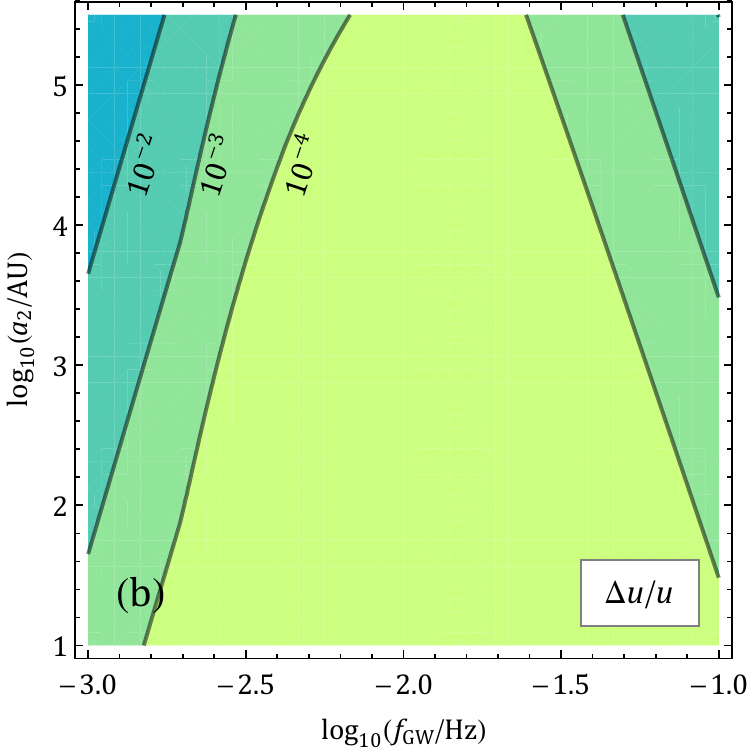}}
\vcenteredhbox{\includegraphics[height=0.32\textwidth]{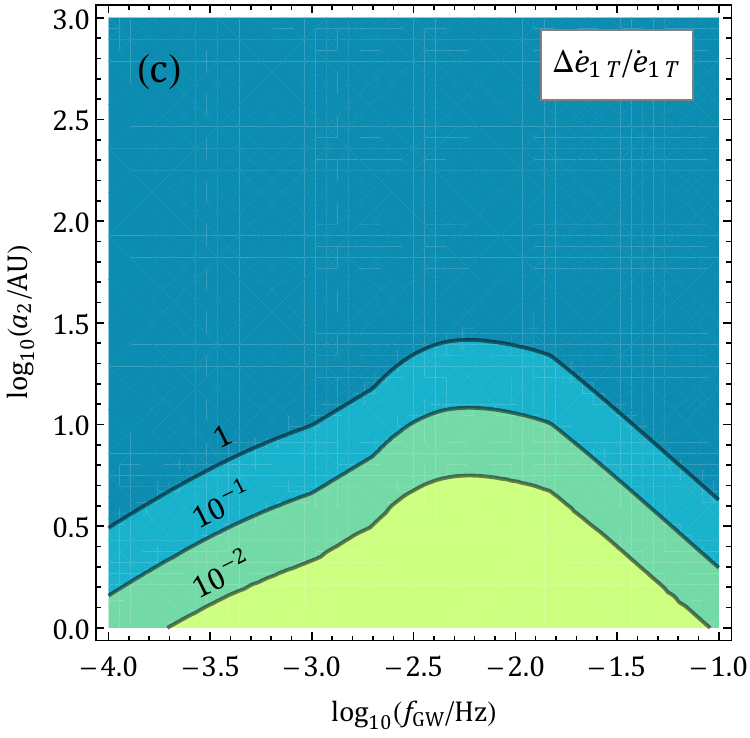}}
\caption{Fisher matrix elements $\Gamma_{\Omega\Omega}^{-1/2}/\Omega$, $\Gamma_{uu}^{-1/2}/u$, and $\Gamma_{\dot e\dot e}^{-1/2}/\dot e_{1T}$, as functions of outer orbital radius $a_2$ and the GW frequency $f_{GW}$. In all plots we take a binary with $m_c=30M_\odot$ at $r=400$Mpc away, orbiting a third body with $m_2=4\times 10^6M_\odot$, observed with $T_O=5$yr. The outer orbit is taken to be edge-on with $I_2=90^\circ$. We take $e_2=0$ in (a,b) and $e_2=0.5$ in (c). Note that these quantities reduce to the relative errors of estimating the parameters $\Omega,u,\dot e_{1T}$, respectively, when the error correlations are weak. (cf. Fig.\;\ref{fig_correlations}).}
\label{fig_errors}
\end{figure*}
 
\begin{figure}[tbph]
\centering
\includegraphics[width=0.48\textwidth]{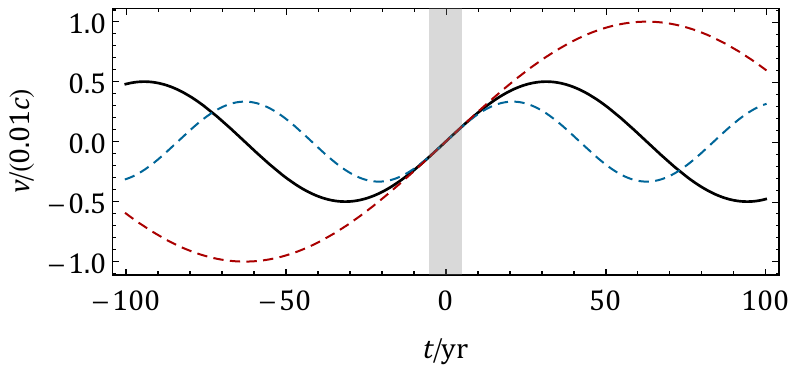}  
\caption{Illustration of the correlation between the outer orbital frequency $\Omega $ and the amplitude of the velocity $u$ with $\Omega T_O\ll 1$. The acceleration $\Omega u$ measured during the observation time $T_O=10$yr (gray region) from the ``true'' $v(t)$ (with $u_*=0.005c$, $\Omega_*=20$yr$^{-1}$) can be fitted, either by larger $\Omega=1.5\Omega_*$ and smaller $u=u_*/1.5$ (blue dashed curve), or by smaller $\Omega=0.5\Omega_*$ and larger $u_*=u/0.5$ (red dashed curve). }
\label{fig_OmegauCorrelation}
\end{figure} 

In Fig.\;\ref{fig_errors}(a) and (b), we take $T_O=5$yr and plot the relative errors $\Delta\Omega/\Omega$ and $\Delta u/u$ as functions of $a_2$ and GW frequency $f_{GW}$ at the beginning of the observation. We take a binary with chirp mass $m_c=30M_\odot$ near a third body with $m_2=4\times 10^6M_\odot$, and the outer orbit is assumed edge-on to maximize the effect. We note that smaller $f_{GW}$ gives larger errors because both the S/N and the phases ($\Phi_\Omega$ and $\Phi_u$) are reduced. On the other hand, larger $f_{GW}$ also gives larger error because the binaries with larger $f_{GW}$ merge faster and thus allow less observation time. In addition, from $u\propto a_2^{-1/2}$ and $\Omega\propto a_2^{-3/2}$, we find the $a_2$-dependence of the errors to be $\Delta\Omega/\Omega\propto a_2^2$ and $\Delta u/u\propto a_2^{1/2}$.

For (\ref{DeltaLambda}) and (\ref{DeltaOmegau}) to hold, we have assumed that the error correlations are small, which is, however, not always true. In fact, two error correlations are of great physical relevance. One is between $\Omega$ and $u$, and can be viewed as a measure of how well we can resolve the outer orbit. Within this two-parameter space,
\begin{align}
&\rh_{\Omega u}^2=\FR{\Gamma_{\Omega u}^2}{\Gamma_{\Omega\Omega}\Gamma_{uu}}\simeq \FR{[\int_0^{T_O}\di t\,(\pd_\Omega \wh h)(\pd_u \wh h)]^2}{[\int_0^{T_O}\di t\,(\pd_\Omega \wh h)^2][\int_0^{T_O}\di t\,(\pd_u \wh h)^2]},\n\\
&\pd_{\Omega}\wh h\simeq -(h_c\cos\Phi) \FR{u}{c} \int_0^{t}\di t' \,f_{GW}(t')\cos(\Omega t'+\ga_2)t',\n\\
&\pd_{u}\wh h\simeq -(h_c\cos\Phi) \FR{1}{c} \int_0^{t}\di t' \,f_{GW}(t')\sin(\Omega t'+\ga_2).
\label{rhoOmegav}
\end{align}
It is intuitively clear that we can measure only a small portion of the whole outer orbit when $a_2$ is so large that $\Omega T_O\ll 1$, in which case the net effect would be a nearly constant acceleration of the binary barycenter. To see this, we Taylor expand the GW phase (\ref{phiobs}) at a given time,
\begin{align}
\label{phiexpand}
  & \Phi(t)= \dot\Phi(0)t+\fr{1}{2}\ddot\Phi(0)t^2+\fr{1}{6}\dddot\Phi(0)t^3+\cdots, \\
  \label{phidot} 
  &\dot\Phi(0)=\big(1-\fr{v_{2\parallel}}{c}\big)f_\text{peak}\big|_{t=0},\\
  \label{phiddot}
  &\ddot\Phi(0)=\big(1-\fr{v_{2\parallel}}{c}\big)\dot f_\text{peak}-\fr{\dot v_{2\parallel}}{c}f_\text{peak}\big|_{t=0}, 
\end{align}
The linear term (\ref{phidot}) is irrelevant because it is degenerate with the redshift. The first nontrivial effect is from (\ref{phiddot}) where $\dot v_{2\parallel}$ is just the barycenter acceleration, which is given by $\sim\Omega u$ for circular orbit. The degeneracy between $\Omega$ and $u$ in this case is illustrated in Fig.\;\ref{fig_OmegauCorrelation}. Quantitatively, this is indicated by a large correlation between $\Omega$ and $u$, namely $\rh_{\Omega u}\sim 1$ as can be seen from (\ref{rhoOmegav}). Oppositely, when $a_2$ is small so that $\Omega T_O\gg 1$, we see that $\rh_{\Omega u}\sim 1/(\Omega T_O)\ll 1$ from (\ref{rhoOmegav}), which means that we have a chance to measure both $\Omega$ and $u$ separately, and thus resolve the whole outer orbit.

\begin{figure*}[t]
\centering
\vcenteredhbox{\includegraphics[height=0.34\textwidth]{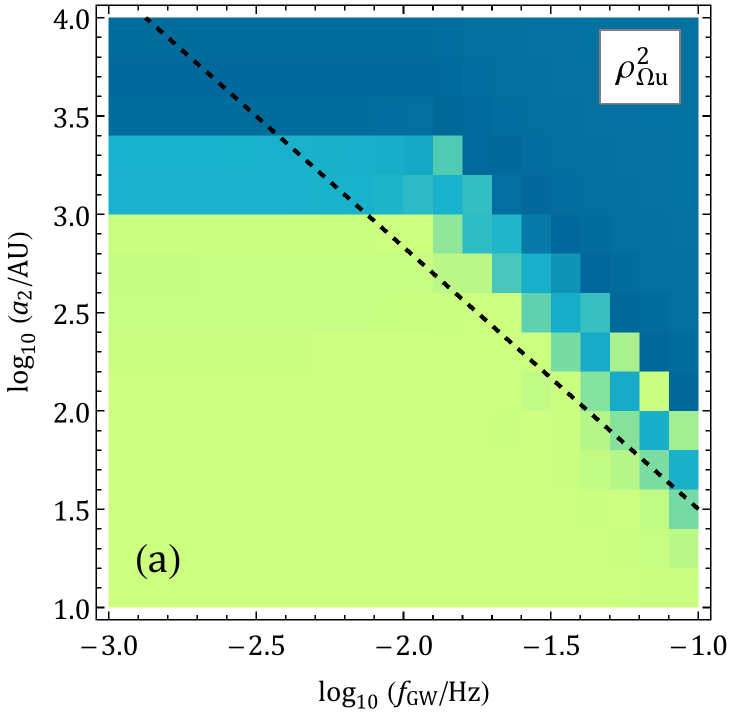}}
\vcenteredhbox{\includegraphics[height=0.34\textwidth]{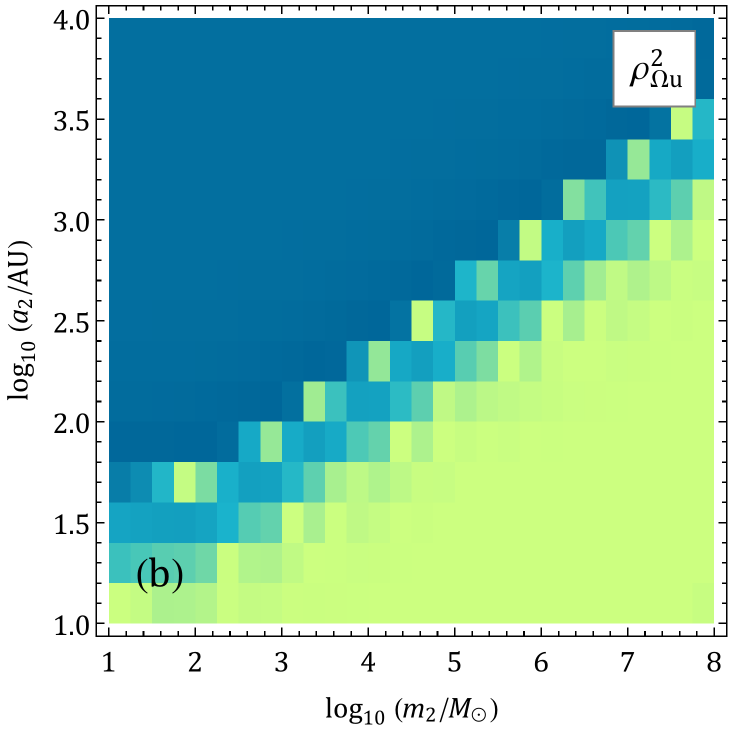}}\\
~~~~~~~~~~~~\vcenteredhbox{\includegraphics[height=0.345\textwidth]{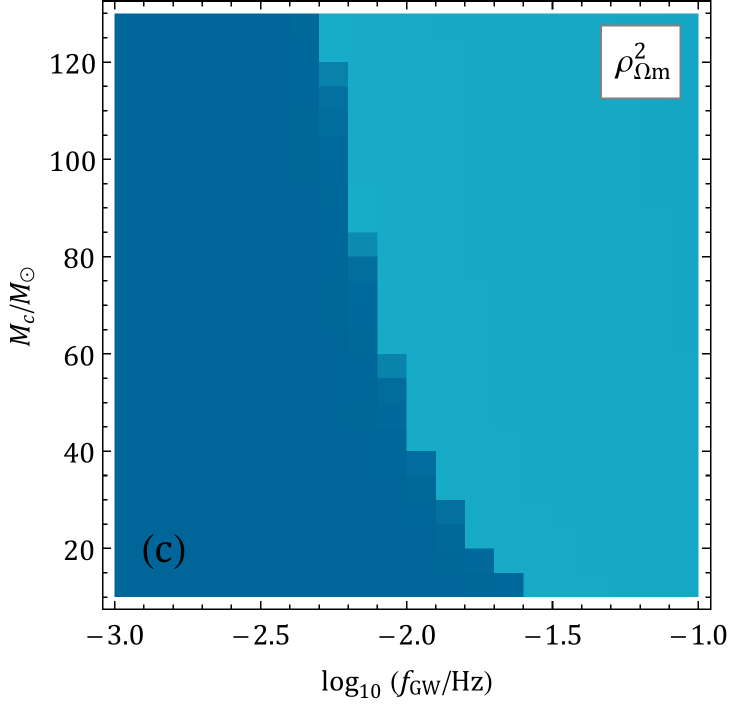}}
\vcenteredhbox{\includegraphics[height=0.36\textwidth]{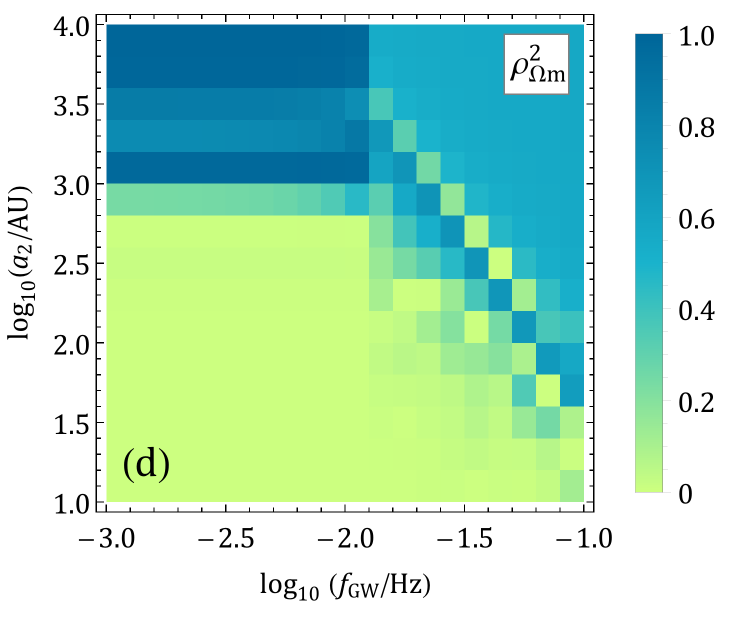}}
\caption{The error correlations $\rh_{\Omega u}^2$ (a, b) and $\rh_{\Omega m}^2$ (c, d). In (a, d), the correlations are plotted as functions of outer orbital radius $a_2$ and the GW frequency $f_{GW}$ when the observation starts. In (b), $\rh_{\Omega u}^2$ is plotted as a function of $a_2$ and $m_2$. In (c), $\rh_{\Omega m}^2$ is plotted as a function of the binary's chirp mass $m_c$ and $f_{GW}$ with $a_2=10^4$AU. We take $m_2=4\times 10^6M_\odot$ in (a,c,d), $m_c=30M_\odot$ in (a,b,d), $f_{GW}=0.01$Hz in (b). In all panels the outer orbit is taken to be edge-on with $I_2=90^\circ$. In (a) we show a dashed line below which there could be ``negatively chirping'' binaries with $\dot f_{GW}<0$.}
\label{fig_correlations}
\end{figure*}

In Fig.\;\ref{fig_correlations}(a) and (b), we plot $\rh_{\Omega u}^2$ and show its dependence on $a_2$, $f_{GW}$, and $m_2$. The strong correlation $\rh_{\Omega u}^2\sim 1$ at large $a_2$ can be clearly seen by noting that $\Omega\simeq\sqrt{GM/a_2^3}$. We also mention the possibility of ``negative chirping'' in which case the phase drift from the outer orbital motion is stronger than the binary's own chirping, resulting in $\dot f_{GW}<0$, so that the outer orbital motion can be unambiguously recognized. In Fig.\;\ref{fig_correlations}(a), ``negative chirping'' could occur below the dashed black line.

Fig.\;\ref{fig_correlations}(b) shows the effect of the third body with different values of $m_2$. It is clear that smaller $m_2$ at a given orbital radius has a smaller effect on the BBH's barycenter motion and therefore the degeneracy between $\Omega$ and $u$ is broken only at smaller $a_2$. For roughly stellar-mass $m_2=\order{10\sim10^2M_\odot}$, visible large orbits require $a_2\lesssim \order{10\sim10^2\text{AU}}$. Coincidentally, this is the typical distance in the field-triple channel \cite{Silsbee:2016djf}. In the case of an SMBH as the third body, $m_2\sim\order{10^{6}\sim 10^8M_\odot}$, large orbits with $a_2\sim\order{10^2\sim 10^3\text{AU}}$ could be visible. Typical merging BBHs in the galactic-center channel live within 0.1pc$\,\sim10^4$AU from the central SMBH. Assuming a mass-segregated density profile for BBHs that yields a number density distribution that is flat in $a_2$, we see that around $\order{1\sim 10\%}$ of BBHs should have $a_2$  small enough to yield visible barycenter motion.

It is worth noting that although $\Omega$ and $u$ cannot be well measured separately for large $a_2$ with $\Omega T_O\ll 1$ due to $\rh_{\Omega u}\sim 1$, the product $\Omega u$ can be measured. This is nothing but the barycenter acceleration of the binary, as considered in \cite{Inayoshi:2017hgw}. But here another error correlation becomes very important, which is between the binary's chirp mass $m_c$ and one of the outer orbital parameter, for which we consider $\Omega$ here as an example. Physically, this correlation $\rh_{\Omega m}^2$ provides a measure of the fact that the Doppler shift from the barycenter motion could be confused by the intrinsic chirping $\dot f_{GW}$, since both of them contribute to $\ddot\Phi$ in (\ref{phiddot}). To break the degeneracy, we need either a significant chirping allowing us to measure $f_{GW}(t)$ beyond the linear level $\dot f_{GW}$, or to observe a significant fraction of the outer orbital motion.  We show how this further restricts the resolvability in Fig.\;\ref{fig_correlations}.

There is a similar correlation between the outer orbital elements ($\Omega$ or $u$) and the GW frequency $f_{GW}(t=0)$ at the initial time of the observation, since increasing $f_{GW}$ would also increase the amount of chirping. We leave a complete analysis of parameter correlations for a future study.

In Fig.\;\ref{fig_correlations}(c), we plot $\rh_{\Omega m}^2$ as a function of binary's chirp mass $m_c$ and the initial GW frequency $f_{GW}(t=0)$ for a binary that is $a_2=10^4$AU away from the third body with $m_2=4\times 10^6M_\odot$. At large distance $a_2=10^4$AU the outer orbital period is $\sim 300$yr so only a small fraction of the outer orbit could be seen. Then Fig.\;\ref{fig_correlations}(c) shows that there is a large correlation between outer orbital motion and the chirping when the chirping is weak, i.e., when both $m_c$ and $f_{GW}$ are small. It is for this reason that \cite{Inayoshi:2017hgw} concluded that the barycenter acceleration cannot be measured for small chirp mass. But here we note that $f_{GW}$, being fixed in the study of \cite{Inayoshi:2017hgw}, is also crucial to determine the strength of chirping and thus the amount of error correlation. In Fig.\;\ref{fig_correlations}(c) we see the strong frequency dependence when determining which range of parameters is resolvable. Fixing the frequency would both under- or overestimate the resolvable regions.

In Fig.\;\ref{fig_correlations}(d), we plot $\rh_{\Omega m}^2$ for different $a_2$ and $f_{GW}(t=0)$ with $m_c=30M_\odot$. We see that the correlation between the chirping and the outer orbital motion is greatly reduced in the near region (small $a_2$) where $\gtrsim1$ outer orbital periods could be observed. Therefore we do not need large chirping to break the degeneracy. \cite{Robson:2018svj} considered these regions assuming little chirping for simplicity. Here we showed that the similar measurement can also be extended to regions with significant chirping and small $a_2$ as long as templates are available. 

\section{Secular Evolution of the Inner Orbital Elements}

We may also observe the secular oscillation of the eccentricity $e_1$ and inclination $I_1$ of the BBH orbit induced by the tidal perturbation of the tertiary body in what is known as the KL oscillation. The tidal perturbation introduces the time variation,
\bge
\label{dedtKL}
  \dot e_{1T}=\FR{15}{8}\sqrt{\FR{Gm_2^2a_1^3e_1^2(1-e_1^2)}{ma_2^6(1-e_2^2)^3}} (1-\cos^2I)\sin 2\ga,
\ede
where $I$ is the inclination of the inner binary and $\ga$ is the argument of the periapsis. Both $I$ and $\ga$ are measured with respect to the outer orbital plane, and 
not with respect to the line of sight.  

The BBH eccentricity $e_1$ might be measured directly from the GW waveform. However, the time scale $\dot e_{1T}$ is  longer than the outer orbital period \cite{Randall:2017jop}, and is generally longer than the observational time $T_O$. Therefore it is difficult to measure directly $e_1(t)$. Instead, the effect of $\dot e_{1T}$ is a net phase drift which could again be confused by other contributions to the phase drift. 

To estimate the region where $\dot e_{1T}$ can possibly be measured, we use the simplified Fisher analysis in last section. To simplify the problem, we assume that $e_1$ is not large so that the GW spectrum is narrow and is dominated by its peak component with frequency in (\ref{fpeak}).  
\begin{align}
\label{dedot}
  &\FR{\Delta\dot e_{1}}{\dot e_{1T}}\simeq \Gamma_{\dot e\dot e}^{-1/2}\simeq \FR{1}{(S/N)|\Phi_{\dot e}|},\n\\
  &\Phi_{\dot e}\simeq \dot e_{1T}(\pd_{e_1}f_\text{peak})T_O^2,
\end{align}
in which we have again assumed that the error correlations are weak, which is not necessarily true for all parameters. We show the relative error (\ref{dedot}) in Fig.\;\ref{fig_errors} for different $a_2$ and $f_{GW}$. We leave a more complete Fisher analysis for future study.

In conjunction with the $e_1$-variation, the BBH inclination $I $ also undergoes secular evolution and there is a corresponding time variation in the observed GW polarization, which is however more challenging to measure.

\section{Expected Number of Detections}

It has been shown in \cite{Sesana:2016ljz} that up to thousands of stellar-mass BBHs could be individually resolved by LISA, given the local merger rate inferred from LIGO detections. On the other hand, the fraction of dynamically formed BBHs is highly uncertain, given the poorly understood merger rates in various dynamical channels. To get a sense of how many of the dynamical BBHs could present measurable orbital motion in LISA, we use the existing literature as a guide to estimate how many events with large-orbital motions might be seen in each channel. By visible large-orbital motion we mean that $\Omega$ and $u$ can both be measured. There could be more binaries with visible phase drift from barycenter acceleration, but with degenerate $\Omega$ and $u$.

A local merger rate $\mathcal{R}_\text{LIGO}\sim 50\text{Gpc}^{-3}\text{yr}^{-1}$ is inferred from the LIGO/Virgo detections. From this \cite{Sesana:2016ljz} showed that up to several thousands of stellar-mass could be visible in LISA, with optimal configuration and running time (N2A5 in 5yrs), although the number could be significantly lower (of order dozens) in a less optimal configuration or in less favorable channels (See, for instance, \cite{Kremer:2018cir,Fang:2019dnh}). Given that the number density is proportional to the merger rate, we can use the fraction of merger rate from dynamical channels to infer the expected fractions of dynamical binaries in LISA. The estimates in the following will be based on the optimal scenario with up to several thousands of resolvable BBHs.

The merger rate of BBHs in galactic centers (with $a_2<0.1$pc) with SMBHs was estimated to be 0.048MWEG$^{-1}$Myr$^{-1}$ \cite{Antonini:2012ad}, which is converted to $0.56$Gpc$^{-3}$yr$^{-1}$ using the conversion MWEG/86Mpc$^3$ \cite{Abbott:2016nhf}. The rate could be reduced if there are not enough BBHs, or if the replenished BBHs are very soft so that they get tidally disrupted at the galactic center. From Fig.\;\ref{fig_correlations} we see that only the innermost BBHs with $a_2<\order{10^3\text{AU}}$ could present visible orbital motion. Comparing this with the typical range $a_2\lesssim 0.1$pc of this channel, and assuming a segregated distribution of BBHs which is flat in $a_2$, we see that up to $\order{10\%}$ of BBHs in this channel could present visible barycenter motion, which could contribute up to several resolvable events in LISA.

There is however the possibility of another channel of BBH mergers in AGNs \cite{Bartos:2016dgn}. The surrounding gases help BBHs to migrate into the galactic center and also provide drag forces that reduce the orbital separation. The rate of merger is estimated to be 12Gpc$^{-3}$yr$^{-1}$. The orbital reduction in this channel is achieved, in addition to GWs, mostly by drag forces as well as Kozai-Lidov resonances. Using the same estimate as above, this channel could contribute up to tens of events with detectable barycenter motions in LISA.

The rate of binaries in globular clusters was predicted to be 14Gpc$^{-3}$yr$^{-1}$ \cite{Rodriguez:2018rmd}. and \cite{Rodriguez:2018pss} showed that $\sim 35\%$ of these BBHs might merge within the cluster instead of earlier ejection, and many of them could show barycenter motion, too. This could contribute again up to tens of events.

The merger rate of BBHs in field triples was predicted to be $\sim$\,6\,Gpc$^{-3}$yr$^{-1}$, with a possible reduction from nonzero natal kicks \cite{Silsbee:2016djf}. Most of the merged binaries in this channel have a small distance to the third body, initially $a_2/a_1\sim\order{1\sim10}$. Therefore we expect that the barycenter motion could be seen for most of BBHs in this channel with preferable orientation of the outer orbit. Therefore this channel may also contribute up to around tens of events in LISA.

\section{Discussion} 

Future space GW telescopes such as LISA will open up a new era of multi-band GW astronomy. 
In this paper we considered the exciting possibility of directly probing the ambient mass surrounding BBHs using LISA through measurements of net phase shifts and also through the observations of waveforms over time,  which will reflect in detail the BBH's orbital motion.  Effects we considered include the barycenter motion of the BBH and the tidal-induced variation of BBH eccentricity. If the BBHs measured by LISA are orbiting around a dense cloud of mass,  such orbital motion can be viewed as a direct measure of  the mass density of the environment. Furthermore the longitudinal velocity fluctuation $\Delta v_{2\parallel}$ can provide meaningful information about the orbital parameters.
 
This information could be critical to distinguishing different formation channels of BBHs. 
Though it would be difficult to 
distinguish globular cluster and galactic center origin apart from eccentricity alone, the BBH barycenter motion would be very different. 
Especially exciting too  is the possibility of following BBHs from LISA into the LIGO band. Even without detailed measurements of the inner orbital parameters at LISA, the outer orbital parameters can be measured from the phase of the waveform alone. This makes it possible to imagine following in detail the waveform from one regime to another, reducing potential degeneracies between inner and orbital parameters further. We leave a more complete analysis, in particular the possible error correlations, for future studies.

\begin{acknowledgments}
We thank Ben Bar-Or, Zolt\'an Haiman, Johan Samsing, and Barak Zackay for helpful conversations, and Neil Cornish for reading the manuscript. LR is supported by an NSF grant PHY-1620806, a Kavli Foundation grant ``Kavli Dream Team,'' and a Simons Foundation grant 511879. LR and ZZX thank the IAS Princeton for their hospitality. 

\end{acknowledgments}


\begin{thebibliography}{}
\expandafter\ifx\csname natexlab\endcsname\relax\def\natexlab#1{#1}\fi
\providecommand{\url}[1]{\href{#1}{#1}}
\providecommand{\dodoi}[1]{doi:~\href{http://doi.org/#1}{\nolinkurl{#1}}}
\providecommand{\doeprint}[1]{\href{http://ascl.net/#1}{\nolinkurl{http://ascl.net/#1}}}
\providecommand{\doarXiv}[1]{\href{https://arxiv.org/abs/#1}{\nolinkurl{https://arxiv.org/abs/#1}}}

\bibitem[{Abbott {et~al.}(2016{\natexlab{a}})}]{Abbott:2016blz}
Abbott, B.~P., {et~al.} 2016{\natexlab{a}}, Phys. Rev. Lett., 116, 061102,
  \dodoi{10.1103/PhysRevLett.116.061102}

\bibitem[{Abbott {et~al.}(2016{\natexlab{b}})}]{Abbott:2016nhf}
---. 2016{\natexlab{b}}, Astrophys. J., 833, L1,
  \dodoi{10.3847/2041-8205/833/1/L1}

\bibitem[{Antonini \& Perets(2012)}]{Antonini:2012ad}
Antonini, F., \& Perets, H.~B. 2012, Astrophys. J., 757, 27,
  \dodoi{10.1088/0004-637X/757/1/27}

\bibitem[{Banerjee(2018)}]{Banerjee:2017mgr}
Banerjee, S. 2018, Mon. Not. Roy. Astron. Soc., 473, 909,
  \dodoi{10.1093/mnras/stx2347}

\bibitem[{Barack \& Cutler(2004)}]{Barack:2003fp}
Barack, L., \& Cutler, C. 2004, Phys. Rev., D69, 082005,
  \dodoi{10.1103/PhysRevD.69.082005}

\bibitem[{Bartos {et~al.}(2017)Bartos, Kocsis, Haiman, \&
  Márka}]{Bartos:2016dgn}
Bartos, I., Kocsis, B., Haiman, Z., \& Márka, S. 2017, Astrophys. J., 835,
  165, \dodoi{10.3847/1538-4357/835/2/165}

\bibitem[{Bonvin {et~al.}(2017)Bonvin, Caprini, Sturani, \&
  Tamanini}]{Bonvin:2016qxr}
Bonvin, C., Caprini, C., Sturani, R., \& Tamanini, N. 2017, Phys. Rev., D95,
  044029, \dodoi{10.1103/PhysRevD.95.044029}

\bibitem[{Breivik {et~al.}(2016)Breivik, Rodriguez, Larson, Kalogera, \&
  Rasio}]{Breivik:2016ddj}
Breivik, K., Rodriguez, C.~L., Larson, S.~L., Kalogera, V., \& Rasio, F.~A.
  2016, Astrophys. J., 830, L18, \dodoi{10.3847/2041-8205/830/1/L18}

\bibitem[{Fang {et~al.}(2019)Fang, Thompson, \& Hirata}]{Fang:2019dnh}
Fang, X., Thompson, T.~A., \& Hirata, C.~M. 2019, Astrophys. J., 875, 75,
  \dodoi{10.3847/1538-4357/ab0e6a}

\bibitem[{{Hamilton} \& {Rafikov}(2019{\natexlab{a}})}]{2019arXiv190201344H}
{Hamilton}, C., \& {Rafikov}, R.~R. 2019{\natexlab{a}}, arXiv e-prints,
  arXiv:1902.01344.
\newblock \doarXiv{1902.01344}

\bibitem[{{Hamilton} \& {Rafikov}(2019{\natexlab{b}})}]{2019arXiv190201345H}
---. 2019{\natexlab{b}}, arXiv e-prints, arXiv:1902.01345.
\newblock \doarXiv{1902.01345}

\bibitem[{Inayoshi {et~al.}(2017)Inayoshi, Tamanini, Caprini, \&
  Haiman}]{Inayoshi:2017hgw}
Inayoshi, K., Tamanini, N., Caprini, C., \& Haiman, Z. 2017, Phys. Rev., D96,
  063014, \dodoi{10.1103/PhysRevD.96.063014}

\bibitem[{Klein {et~al.}(2016)}]{Klein:2015hvg}
Klein, A., {et~al.} 2016, Phys. Rev., D93, 024003,
  \dodoi{10.1103/PhysRevD.93.024003}

\bibitem[{Kozai(1962)}]{Kozai:1962zz}
Kozai, Y. 1962, Astron. J., 67, 591, \dodoi{10.1086/108790}

\bibitem[{Kremer {et~al.}(2019)}]{Kremer:2018cir}
Kremer, K., {et~al.} 2019, Phys. Rev., D99, 063003,
  \dodoi{10.1103/PhysRevD.99.063003}

\bibitem[{Lidov \& Ziglin(1976)}]{Lidov:1976qhg}
Lidov, M.~L., \& Ziglin, S.~L. 1976, Celestial Mech., 13, 471,
  \dodoi{10.1007/BF01229100}

\bibitem[{Meiron {et~al.}(2017)Meiron, Kocsis, \& Loeb}]{Meiron:2016ipr}
Meiron, Y., Kocsis, B., \& Loeb, A. 2017, Astrophys. J., 834, 200,
  \dodoi{10.3847/1538-4357/834/2/200}

\bibitem[{Nishizawa {et~al.}(2016)Nishizawa, Berti, Klein, \&
  Sesana}]{Nishizawa:2016jji}
Nishizawa, A., Berti, E., Klein, A., \& Sesana, A. 2016, Phys. Rev., D94,
  064020, \dodoi{10.1103/PhysRevD.94.064020}

\bibitem[{Nishizawa {et~al.}(2017)Nishizawa, Sesana, Berti, \&
  Klein}]{Nishizawa:2016eza}
Nishizawa, A., Sesana, A., Berti, E., \& Klein, A. 2017, Mon. Not. Roy. Astron.
  Soc., 465, 4375, \dodoi{10.1093/mnras/stw2993}

\bibitem[{Randall \& Xianyu(2018{\natexlab{a}})}]{Randall:2017jop}
Randall, L., \& Xianyu, Z.-Z. 2018{\natexlab{a}}, Astrophys. J., 853, 93,
  \dodoi{10.3847/1538-4357/aaa1a2}

\bibitem[{Randall \& Xianyu(2018{\natexlab{b}})}]{Randall:2018nud}
---. 2018{\natexlab{b}}, Astrophys. J., 864, 134,
  \dodoi{10.3847/1538-4357/aad7fe}

\bibitem[{Robson {et~al.}(2018)Robson, Cornish, Tamanini, \&
  Toonen}]{Robson:2018svj}
Robson, T., Cornish, N.~J., Tamanini, N., \& Toonen, S. 2018, Phys. Rev., D98,
  064012, \dodoi{10.1103/PhysRevD.98.064012}

\bibitem[{Rodriguez {et~al.}(2018)Rodriguez, Amaro-Seoane, Chatterjee, Kremer,
  Rasio, Samsing, Ye, \& Zevin}]{Rodriguez:2018pss}
Rodriguez, C.~L., Amaro-Seoane, P., Chatterjee, S., {et~al.} 2018, Phys. Rev.,
  D98, 123005, \dodoi{10.1103/PhysRevD.98.123005}

\bibitem[{Rodriguez \& Loeb(2018)}]{Rodriguez:2018rmd}
Rodriguez, C.~L., \& Loeb, A. 2018, Astrophys. J., 866, L5,
  \dodoi{10.3847/2041-8213/aae377}

\bibitem[{Rodriguez {et~al.}(2016)Rodriguez, Zevin, Pankow, Kalogera, \&
  Rasio}]{Rodriguez:2016vmx}
Rodriguez, C.~L., Zevin, M., Pankow, C., Kalogera, V., \& Rasio, F.~A. 2016,
  Astrophys. J., 832, L2, \dodoi{10.3847/2041-8205/832/1/L2}

\bibitem[{Samsing \& D'Orazio(2018)}]{Samsing:2018isx}
Samsing, J., \& D'Orazio, D.~J. 2018, Mon. Not. Roy. Astron. Soc., 481, 5445,
  \dodoi{10.1093/mnras/sty2334}

\bibitem[{Sesana(2016)}]{Sesana:2016ljz}
Sesana, A. 2016, Phys. Rev. Lett., 116, 231102,
  \dodoi{10.1103/PhysRevLett.116.231102}

\bibitem[{Silsbee \& Tremaine(2017)}]{Silsbee:2016djf}
Silsbee, K., \& Tremaine, S. 2017, Astrophys. J., 836, 39,
  \dodoi{10.3847/1538-4357/aa5729}

\bibitem[{Wen(2003)}]{Wen:2002km}
Wen, L. 2003, Astrophys. J., 598, 419, \dodoi{10.1086/378794}

\bibitem[{Yunes {et~al.}(2011)Yunes, Coleman~Miller, \&
  Thornburg}]{Yunes:2010sm}
Yunes, N., Coleman~Miller, M., \& Thornburg, J. 2011, Phys. Rev., D83, 044030,
  \dodoi{10.1103/PhysRevD.83.044030}

\end{thebibliography}
\end{document}